\documentclass[sigconf]{acmart}
\renewcommand\footnotetextcopyrightpermission[1]{} 
\usepackage{algorithmic}
\usepackage{graphicx}
\usepackage{textcomp}
\usepackage{xcolor}
\usepackage{listings}

\usepackage{color,soul}
\usepackage{fancyhdr}
\usepackage[normalem]{ulem}

\usepackage{comment}
\usepackage{tabularx}
\usepackage{booktabs}
\usepackage{amsbsy}
\usepackage{pifont}
\usepackage{xspace}

\newcommand{\proj}{wsFFT\xspace}
\newcommand{\wsfft}{wsFFT\xspace}

\newcommand{\device}{WSE\xspace}
\newcommand{\ws}{\device}

\AtBeginDocument{%
  \providecommand\BibTeX{{%
    \normalfont B\kern-0.5em{\scshape i\kern-0.25em b}\kern-0.8em\TeX}}}

\begin{document}

\author{Marcelo Orenes-Vera}
\affiliation{
  \institution{Princeton University}
  \city{Princeton}
  \country{USA}
}
\email{movera@princeton.edu}

\author{Ilya Sharapov}
\affiliation{
  \institution{Cerebras Systems}
  \city{Sunnyvale}
  \country{USA}
}
\email{ilya@cerebras.net}

\author{Robert Schreiber}
\affiliation{
  \institution{Cerebras Systems}
    \city{Sunnyvale}
  \country{USA}
}
\author{Mathias Jacquelin}
\affiliation{
  \institution{Cerebras Systems}
    \city{Sunnyvale}
  \country{USA}
}
\author{Philippe Vandermersch}
\affiliation{
  \institution{Cerebras Systems}
    \city{Sunnyvale}
  \country{USA}
}
\author{Sharan Chetlur}
\affiliation{
  \institution{Cerebras Systems}
    \city{Sunnyvale}
  \country{USA}
}

\renewcommand{\shortauthors}{Marcelo Orenes-Vera, et al.}

\title{Wafer-Scale Fast Fourier Transforms}

\begin{abstract}

We have implemented fast Fourier transforms for one, two, and three-dimensional arrays on the Cerebras CS-2, a system whose memory and processing elements reside on a single silicon wafer.
The wafer-scale engine (\ws) encompasses a two-dimensional mesh of roughly 850,000 processing elements (PEs) with fast local memory and equally fast nearest-neighbor interconnections.

Our wafer-scale FFT (\wsfft) parallelizes a $n^3$ problem with up to $n^2$ PEs.
At this point a PE processes only a single vector of the 3D domain (known as a pencil) per superstep, where each of the three supersteps performs FFT along one of the three axes of the input array.
Between supersteps, \wsfft redistributes (transposes) the data to bring all elements of each one-dimensional pencil being transformed into the memory of a single PE.
Each redistribution causes an all-to-all communication along one of the mesh dimensions.
Given the level of parallelism, the size of the messages transmitted between pairs of PEs can be as small as a single word.
In theory, a mesh is not ideal for all-to-all communication due to its limited bisection bandwidth.
However, the mesh interconnecting PEs on the \ws lies entirely on-wafer and achieves nearly peak bandwidth even with tiny messages.

This high efficiency on fine-grain communication allow \wsfft to achieve unprecedented levels of parallelism and performance.
We analyse in detail computation and communication time, as well as the weak and strong scaling, using both FP16 and FP32 precision.
With 32-bit arithmetic on the CS-2, we achieve 959 microseconds for 3D FFT of a $512^3$ complex input array using a $512\times512$ subgrid of the on-wafer PEs.
This is the largest ever parallelization for this problem size and the first implementation that breaks the millisecond barrier.

\end{abstract}


\maketitle
\thispagestyle{plain}
\pagestyle{plain}

\vspace{-1mm}
\section{Introduction}\label{sec:intro}

The discrete Fourier transform (DFT) is a cornerstone of scientific computing. It lies at the heart of many scientific applications — in 
signal and image processing,
differential equations, 
number theory, 
cryptography,
probability and statistics, 
and other areas.
The most common way to compute the DFT is to use one of the many formulations of the Fast Fourier Transform (FFT).
For a $k$-dimensional dataset of size $n^k$ (complex numbers), it takes $k \ n^{k-1} 5 n \log_2 n$ real arithmetic operations~\cite{vanloan}.
(The data are repeatedly transformed along each dimension, and the dimensions are taken in any order. We call these \textit{supersteps}.)

In the context of large-scale scientific simulation or signal analysis, one or more FFTs may be computed at every timestep.
It is therefore crucial that this operation achieves optimal performance on parallel platforms.
We explore multidimensional FFTs using the Cooley-Tukey algorithm~\cite{cooley_tukey}, which is widely used for its efficiency, but whose parallel implementations face several \textbf{challenges}:

\begin{itemize}
    \item The inner loop accesses data with varying, non-unit strides.
    \item For each of the $k$ supersteps, the transformed dimension is typically mapped to memory while the other dimensions are distributed across the PEs. This calls for a redistribution of the data, or transpose, after each superstep.
    \item The redistribution is a nonlocal, all-to-all communication along one-dimensional subsets of the processor grid. For some problem sizes, this limits performance due to the small bisection bandwidth of a grid.
    \item The fine-grained parallelization that we employ leads to fewer data per node and thus smaller messages. This makes both communication latency and overhead of message send or receive more significant and would hurt the efficiency of inter-node communication on coarse-grained architectures using MPI or other messaging software.
\end{itemize}

Parallel computation of the FFT on shared and conventional distributed memory systems is well-understood~\cite{takahashi2003_openmp}.
However, when scaling to large clusters, performance and code complexity are limited by issues related to the use of many single-chip multiprocessors: complex memory hierarchies, heterogeneous compute nodes that marry a CPU to one or several GPUs, and the need to use message passing for inter-node communication when transposing in the multidimensional case~\cite{takahashi2009_cluster,dalcin_advanced_mpi,heffte,p3dfft,pfft}.

The Wafer-Scale Engine (\ws) and the CS-1 system introduced an unprecedented level of parallelism within a single chip---two orders of magnitude more processing elements (PEs) than state-of-the-art GPUs.
The CS-2 system is based on a 7nm generation of the \ws, a $46,225 mm^2$ silicon wafer containing all the system's processing and memory---see~Section~\ref{sec:wse}.
The \ws has a private memory per PE---removing the need for caches---and an interconnection network whose injection bandwidth equals the compute bandwidth, even for single-word messages.
The CS-2 occupies a third of a standard datacenter rack and draws 20 KW of power.

\textbf{Our approach:}
We explore using the Cerebras \ws to compute FFTs and address questions regarding program complexity, single-PE performance, communication for data redistribution, and memory capacity.
Our wafer-scale FFT (\proj{}) leverages the (nearly million-fold) parallelism and the 40GB memory capacity of the \ws, to parallelizes a 3D FFT of size $n^3$ with $O(n^3 \log_2 n)$ operations with $n^2$ PEs in $O(n^2)$ time. 
This allows us to transform in memory a $512^3$ array using $512\times512$ PEs in under a millisecond.
Our 3D FFT comprises three supersteps, one per dimension of the input array.
During each of these, every PE holds and transforms a single pencil (one-dimensional data subset).
This involves having two dimensions of the 3D input array mapped to the 2D mesh, and the third in memory, that is, input$(i,j,k)$ is initially stored as element $k$ of the pencil on PE $(i,j)$.
Section~\ref{sec:how} elaborates on how we efficiently compute within each PE and communicate across PEs.

We start the results section by reporting single-PE performance (Section~\ref{sec:pencil_compute_results}).
In Section~\ref{sec:breakdown}, we report \proj{}’s performance on CS-2 for problems of size $n^3$, mapped into a fabric of $n^2$ PEs, where $n$ ranges from $32$ to $512$.
In Section~\ref{sec:scaling}, we study strong scaling by mapping a given problem on a smaller subset of the CS-2, e.g., $1024^3$ FFT using $512^2$ PEs.
This requires each PE to compute and communicate more than one pencil during data redistributions, which is not currently implemented.
However, after analyzing the theoretical and experimental performance of \proj{}, we provide upper-bound runtime for these strong scaling scenarios.
In Section~\ref{sec:comparison}, our measured datapoints and out scaling models show that the achieved performance is comparable to supercomputers for problems up to the CS-2 memory capacity in multidimensional situations.
Furthermore, \proj is 18\% faster than the fastest reported FFT of size $512^3$~\cite{nvidia_fft}, with the same precision.
Unsurprisingly, due to the proximity of the computing elements and fast interconnect, \proj{} is an order of magnitude faster than best-performing evaluations on CPU clusters~\cite{heffte,plimpton, fft_on_mira}.

\paragraph{Memory capacity}
The storage requirement of the out-of-place, complex-to-complex 3D FFT, at 8-bytes-per datum (a complex number using FP32) is $2 \times 8 \ n^3$ bytes. 
Thus, the largest power-of-two size of $n$ that fits within the \ws memory is $n = 1024$, which takes $2^{34} = 16GiB$.
For 2D data, $n = 2^{15}$ would be possible.
These size ranges seem adequate as they are used often for applications such as molecular dynamics~\cite{lammps}.

\paragraph{Data placement}
We have considered in situ transforms in which the data are present on the \ws before the FFT, as the Fourier transform is a common workload used within larger applications (like the ones mentioned at the beginning of this section).
We have not investigated using the \ws as an attached accelerator for FFT, in which inputs and outputs are transferred into and out before and after the computation.
The time for that data movement would dominate in such a use case.
The maximum transfer bandwidth to the wafer is 1.2 Terabits/s for the current generation of the \ws, so transferring a $512^3$ array would take at least 14 ms.
This contrasts with the $\sim$1 ms of execution of such problem size.


In the following sections, we provide background on multidimensional FFT parallelization and describe the CS-2 and the \ws.
We then describe the \proj{} approach, give measured performance results, and finally discuss prior FFT implementations and other related topics.

\textbf{Our technical contributions are}:
\begin{itemize}
    \item Implementing multidimensional FFTs on a distributed memory architecture using hundreds of thousands of computing elements on a single silicon die while displaying good strong scaling properties.
    \item The largest parallelization of 3D FFT achieving speedups even when each PE only processes one pencil per superstep.
    \item Detailed analysis of the computation and communication time of FFT on the \ws, with FP16 and FP32 precision types.
\end{itemize}

\textbf{We evaluate and demonstrate:}
\begin{itemize}
    \item The fastest-ever FFT for problem sizes that fit within the storage capacity of the CS-2. 
    \item Experimental performance that closely matches theoretical performance for individual pencil computation, leveraging the per-PE SIMD capabilities.
    \item Highly predictable runtime of FFT, as all the data is local for computations and communication is deterministic. This is demonstrated experimentally: several iterations of forward and inverse FFT loops take nearly identical times.
\end{itemize}

\section{Background on FFT Parallelization}\label{sec:background}

Parallel formulations of multidimensional FFT require communication between PEs.
Communication is usually the speed-limiting factor.
Because of its ubiquitous utility and heavy communication, FFT is an excellent stress test for the communication system of a parallel machine, hence its inclusion in the NAS Parallel Benchmarks~\cite{nas_benchmark}.
We restrict our attention to distributed memory implementations.

\subsection{Data Distribution}
The first question about any distributed-memory approach is the distribution of the input and output data to the PEs.

\paragraph{Slab decomposition} This strategy distributes a single axis of a multidimensional array onto a parallel platform~\cite{dalcin_advanced_mpi}. For example, on a 3D problem, it assigns a matrix of elements to each processing element, i.e., input$(i,j,k)$ is stored as a 2D array in the memory of PE$(k)$.
This strategy leverages the efficiency of coarse-grain communication, but it is, by definition, limited to a number of PEs equal to the number of elements of the longest dimension.

\paragraph{Pencil decomposition}
In this approach, the input data is viewed as a 2D array of 1D pencils, and each PE stores a small subarray of the 2D array of such pencils.
In each of the three supersteps, the PEs perform 1D FFT on the pencils they store without communication.
Then between supersteps, the PEs communicate to change which axis of the data is stored in memory.
This achieves higher levels of parallelism since two axes of the 3D data array are distributed.
This is the approach taken by Takahashi~\cite{takahashi2009_cluster} to parallelize to up to 4096 processors.
In this paper, we choose this decomposition for its parallelism and the ease of mapping to the 2D mesh of the \ws. 

\subsection{Phases of the FFT}
For simplicity, in the remainder of the section, we consider the example of a $256^2$ array of axes $i$ and $j$, such as the pixel array of a monochrome image.
The FFT of a 2D array is usually done in two stages.
First, each row $j$ of the image (a vector of length 256) is transformed using a 1D FFT.
The most straightforward parallel strategy is to allocate each row to a single PE, using a total of 256 PEs in this case.
This first stage involves only local computation; there is no communication, and the workload is perfectly balanced.

Following this first local compute phase (or \textit{superstep}), it needs to compute 1D FFTs of each of the 256 pencils along the $j$ axis.
Normally, one of two methods is used:
(1) (transpose algorithm) A global redistribution of the multidimensional array that brings the data needed for one or several pencils into the memory of each PE, as done, e.g., by Gupta and Kumar~\cite{kumar_decom}, or
(2) (distributed transforms) Performing the second superstep with the data in place by exchanging pairs of array elements between pairs of PEs during the transformation of the distributed pencils that lie in the $j$ axis.

The second method has a more complicated communication pattern (butterfly), which is intrinsic to the Cooley-Tukey radix rearrangement~\cite{cooley_tukey}.
In this work, we use the transpose algorithm, which has performed better for large sizes in earlier work~\cite{pencil_decom}.
Gupta and Kumar~\cite{kumar_decom}, and Foster and Worley~\cite{worley_decom} review both methods.

To formalize it using our 2D image example, the input data is $a(i,j)$, $0 <= i,j < 256$, where $a(i,j)$ is stored in PE$(j)$.
Once each PE computes an FFT over its local pencil, the first superstep results in y = 1D\_FFT$(a)$ where $y(i,j)$ is also in PE$(j)$.
The data communication then transposes $y$ so that $y(i,j)$ is stored in PE$(i)$, allowing all the PEs to compute the second superstep locally again.


Extending pencil decomposition to a higher number of dimensions may seem intuitive by now: 
a 3D FFT can, for instance, be performed as a sequence of single-pencil FFTs along each dimension.
For example, the FFT of an $N_x$ × $N_y$ × $N_z$ array can be accomplished by a first superstep of transforming $N_x$ × $N_y$ pencils of size $N_z$, followed by two other supersteps along the other two axes.
The \ws is a 2D-mesh with more than $512$ PEs per dimension, and this pencil decomposition unleashes its full computing power as each pencil can be computed in parallel during a superstep.

\section{The wafer-scale engine (\ws)}\label{sec:wse}

The \ws is the world’s largest silicon chip.
It is fabricated on the 215mm-square center of a 300mm-diameter silicon wafer.
The first version of this chip (installed in the CS-1 system) had nearly 400,000 processing elements, while the \ws-2 (installed in the CS-2 system) has more than twice as many.
In this paper, we focus on the \ws-2.
For fine-grained, highly parallel computations, wafer-scale integration dramatically reduces the amount of off-chip communication compared to implementations that span ensembles of processors.

The wafer is organized as a 2D mesh, referred to as \textit{fabric}, connecting around 850,000 processing elements (PEs).
Each PE contains a compute core, a 48KB private memory, and a router.
Each router has bidirectional links to its four neighbors, as well as a full-duplex link to the compute core.
The core has a set of input and output queues, providing an asynchronous interface with its router.


\subsection{Programming Model}
The \ws is a distributed memory system where a PE may access only its own memory.
There is no cache: the local memory supplies all operands fast enough to saturate the compute capacity.
There is no off-chip memory and no DRAM.
The compute core offers a fused multiply-accumulate (FMAC) instruction on 32-bit floating-point numbers, one per clock.
It also supports SIMD instructions: for instance, the SIMD FMAC can perform four 16-bit fused multiply-add floating-point operations per cycle.

The \ws relies on a task-based programming model in which short tasks are often triggered by incoming messages.
A task executes to completion; then, the hardware picks the next task to execute among the ones ready to run.
Tasks are often lightweight: a task may handle the arrival of a single scalar operand from the fabric, for example, and frequently does so with an instruction that may operate on in-memory vectors along with the arriving scalar.

Instructions generally execute and complete in order. 
However, when an instruction over array data involves a vector operand coming from or going to the fabric, it can be run as a background thread (called a \textit{microthread}).
This allows each core in the \ws to perform computations while waiting for incoming/outgoing data.
The microthreaded instruction will make progress when data appears in the input queues or when free slots become available in the output queues.
A common practice is making a handler task eligible to be run only after the completion of a microthread, to enforce a flow dependence or antidependence between two tasks.

In the Cerebras ISA, performance features of the architecture such as the data-triggered task model, microthreads, vector instructions, and router programming, are exposed to the programmer to optimize communication and synchronization.
The same code can be shared by groups of PEs, and it is broadcast to the PEs before the start of the program.
The data is distributed across the PEs according to the scheme specified in the SDK's Python binding.


\begin{figure}[t]
\centering  
\includegraphics[width=\columnwidth]{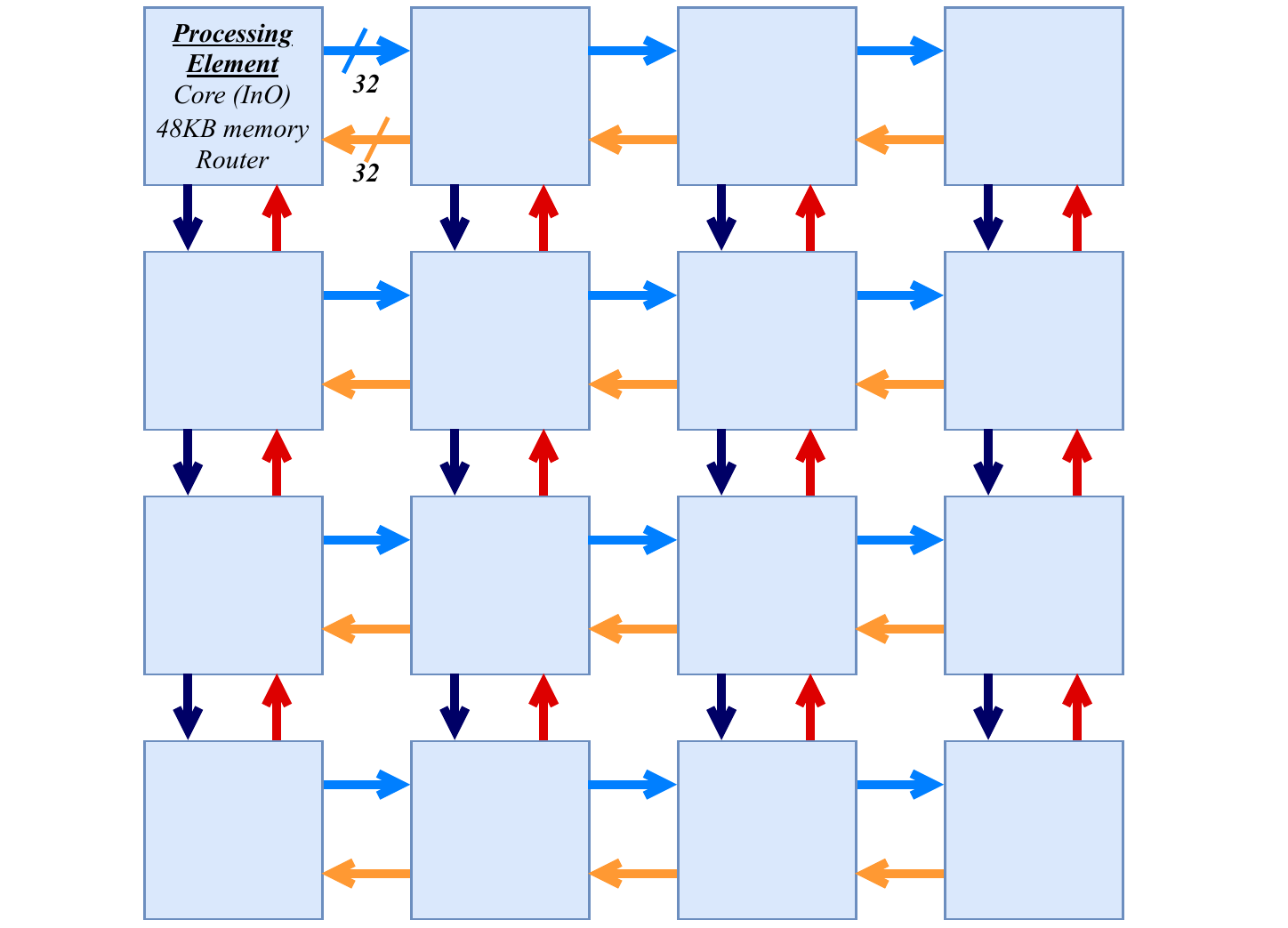}
\vspace{-1mm}
\caption{A subset of $4 \times 4$ PEs within the \ws.
The links between PEs are 32-bit wide and bi-directional, thus, a word can be transferred in each direction in parallel for each cycle.
Each PE contains a 48KB memory, an in-order core, and a router to connect to its four neighbors and to the core. 
}
\vspace{-7mm}
\label{fig:mesh}
\end{figure}

\subsection{On-chip Communication}
Figure~\ref{fig:mesh} shows a subset of $4 \times 4$ PEs.
The bandwidth of a link is a 32-bit word per clock, and the latency of a hop is one clock cycle on average. 
Communication instructions (send, receive) are built into the Cerebras instruction set.
PEs communicate in streams of 32-bit packets that the \ws refers to as wavelets.
A software message can be as small as one hardware wavelet.

Routing wavelets is done at the hardware level: streams of wavelets are routed along logical channels called \textit{colors}.
There are 24 of them available to the programmer.
Every router has a programmable routing table indicating, for every color, which subset of the five outgoing links (four neighbors and the local core) are to receive an incoming wavelet of a given color.
Routing tables can be changed dynamically.
The fabric uses dedicated input buffers for each color at each router and ensures that no wavelet is dropped and that the communication remains deadlock-free.
Only one incoming link can send wavelets of a given color, i.e., there can never be a collision of two incoming wavelets vying for a buffer slot of the same color.

The limited number of colors means that for some applications, including FFT, it is impossible to dedicate a color to every pair of communicating PEs.
However, a single color can be used by adding a filtering option that lets the router select specific wavelets to be routed to the core while allowing other wavelets through.

To scatter data across PEs, it is desirable to send the entire data sequence along a row or a column of PEs and use count-based filters so that each router captures and sends to the local compute core a specific subset of the wavelets that flow on a particular color.


When a wavelet is routed to a compute core, it is put in an input queue.
A PE can accept incoming wavelets in two ways.
\begin{itemize}
    \item A single wavelet can trigger the execution of a handler task,
    \item A stream of incoming wavelets can be treated as a one-dimensional vector operand of a microthreaded instruction.
\end{itemize}

We use the latter option, together with router filters, to implement and optimize the all-to-all communication needed for transposing data at the end of each superstep.

\section{The Approach: Wafer-Scale FFT}\label{sec:how}

\definecolor{codegreen}{rgb}{0,0.6,0}
\definecolor{codegray}{rgb}{0.5,0.5,0.5}
\definecolor{codepurple}{rgb}{0.58,0,0.82}
\definecolor{backcolour}{rgb}{0.95,0.95,0.92}
\lstdefinestyle{mystyle}{
    backgroundcolor=\color{backcolour},   
    commentstyle=\color{codegreen},
    keywordstyle=\color{blue},
    numberstyle=\tiny\color{codegray},
    stringstyle=\color{codepurple},
    basicstyle=\ttfamily\footnotesize,
    breakatwhitespace=false,         
    breaklines=true,                 
    captionpos=b,                    
    keepspaces=true,                 
    numbers=left,                    
    numbersep=5pt,                  
    showspaces=false,                
    showstringspaces=false,
    showtabs=false,                  
    tabsize=2,
    morekeywords={array, operand, function, task}
}

\lstset{style=mystyle}
\begin{lstlisting}[language=C, caption={Pseudocode that computes the FFT of a pencil.}, label=lst:pencil_code, style=mystyle,float]
param N
uint16 N_half = N / 2
float32 x[N*2]
float32 aux[N]
float32 roots_of_unity[N]

// Initialize the descriptors of the array operations
array operand x_even = {.base=x, .length=N}
array operand x_odd  = {.base=x+N, .length=N}
array operand x_aux   = {.base=aux, .length=N}

function reshape (base_dst, base_src, stride_src){
  uint16 stride_dst = stride_src /2
  while(baseB < N){
    array operand x_dst = {.base=x+base_dst, .length=base_src}
    array operand x_src = {.base=x+base_src, .length=base_src}  
    x_dst[] = x_src[]  
    base_src += stride_src 
    base_dst += stride_dst
  }
}

task fft {
 uint16 subproblems = N
 while (subproblems > 1) {
  uint16 offset = 0
  
  while (offset < N){
    float32 factor_real = roots_of_unity[offset]
    float32 factor_imag = roots_of_unity[offset+1]
    // Applying the roots of unity to the odd subarray
    array operand x_odd_real = {.base=x+N+offset,                  .length=subproblems, .stride=2 }
    array operand x_odd_imag = {.base=x+N+offset+1,                .length=subproblems, .stride=2 }
    array operand aux_offset_real = {.base=aux+offset,             .length=subproblems, .stride=2 }
    array operand aux_offset_imag = {.base=aux+offset+1,           .length=subproblems, .stride=2 }
    
    // Complex-number multiplications are broken down into Real Arithmetic 'Add' and 'Mul'.
    aux_offset_real[] = x_odd_real[] * factor_real
    aux_offset_imag[] = x_odd_imag[] * factor_real
    // Operations leveraging native FMAC support
    aux_offset_real[] += x_odd_imag[] * factor_imag
    aux_offset_imag[] += x_odd_real[] * -factor_imag
    offset += subproblems
  }
  
  // Applying even and odd sub-arrays.
  x_odd[]  = x_even[] - x_aux[]
  x_even[] = x_even[] + x_aux[]
  x_aux[]  = x_odd[]
  
  // We re-shape the array to keep the even and odd array elements contiguous
  if (subproblems > 2){
    //Fill Right part of the array
    reshape(N,       subproblems, subproblems)
    reshape(N+N_half,subproblems, subproblems)
    //Fill Left part of the array
    reshape(0,       0,  subproblems)
    reshape(N_half,  0,  subproblems)
  }
  subproblems >>= 1;
 }
}
\end{lstlisting}

\begin{figure*}[t]
\centering  
\vspace{-2mm}
\includegraphics[width=\textwidth]{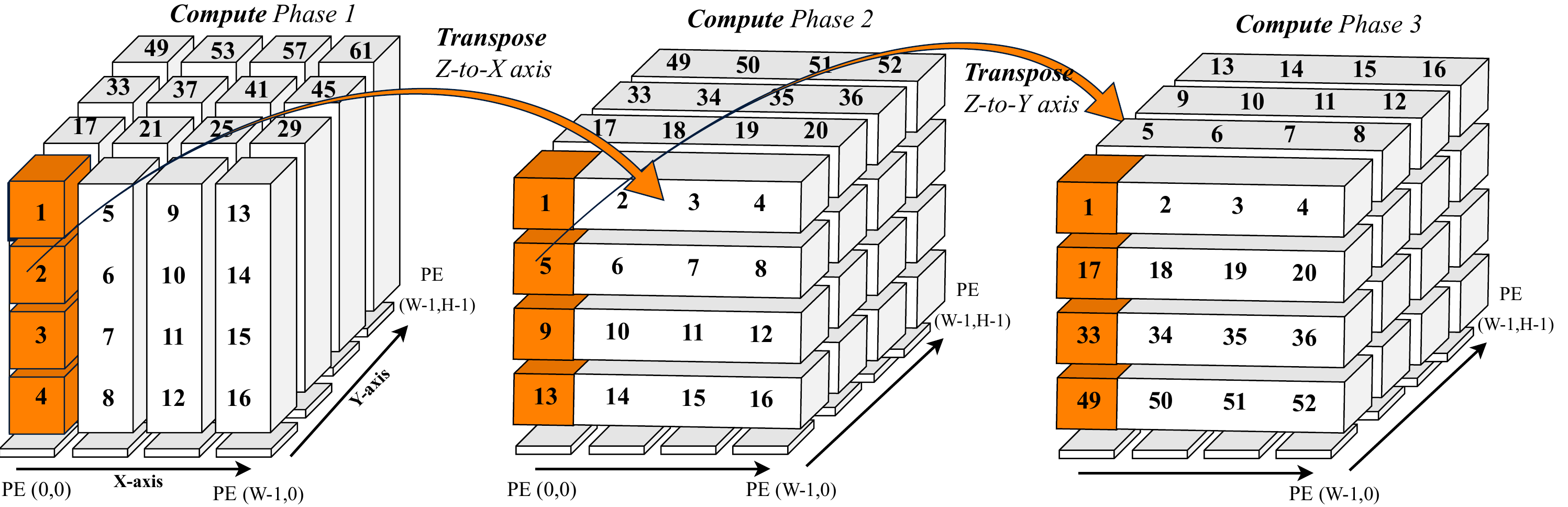}
\vspace{-4mm}
\caption{Mapping of \proj{} for a 3D example problem of size $4^3$ into a 4 PE by 4 PE region of the compute fabric. 
The aforementioned pencil decomposition requires computing $4^2$ pencils of size $n=4$ per dimension, with pencils computed as 1D arrays along each axis.
A transpose phase is necessary between computing phases to keep pencil computation local to a PE.
The first phase calculates data along the z-axis, and the transposition places the data accordingly so that in subsequent phases compute FFTs over the other axes.
The computing phases are regular thanks to the transpositions, and so each pencil is always mapped to a single PE, e.g., the pencil colored in orange is computed on the PE with coordinates x=0 and y=0.
}
\vspace{-1mm}
\label{fig:problem_mapping}
\end{figure*}

Multidimensional FFTs can be decomposed as a sequence of lower-dimensional FFTs.
As mentioned before, typical approaches in parallel processing are \textit{slab} decomposition (2D arrays) or \textit{pencil} decomposition (1D arrays).
Processing the 3D FFT as a matrix of pencils matches well to the 2D mesh of Processing Elements (PE) of the \ws, to maximize parallelism.

The \proj{} code is homogeneous to all PEs, and only the PE coordinates (needed for the transpose stage) are mapped differently---these coordinates indicate the X and Y position of PEs within the rectangular submesh in which the application runs.
Then, each slice of the input array is sent to the corresponding PE, which triggers the start of the program.
Section~\ref{sec:3d_mapping_approach} details the distribution of the input array across the PEs.

\subsection{Pencil Computation}\label{sec:pencil_compute_approach}

The pseudocode in Listing~\ref{lst:pencil_code} concisely describes the part of the per-PE code that computes the FFT of a pencil.
We validated that both the 1D FFT described here and the multidimensional FFTs implemented in this paper are correct by comparing the outputs with those obtained from the Numpy implementation.\footnote{https://numpy.org/doc/stable/reference/routines.fft.html}

The \texttt{fft task} of Listing~\ref{lst:pencil_code} performs a complex-to-complex Fourier transformation, following the Cooley-Tukey algorithm. 

In the radix-2 formulation of Cooley-Tukey, the input data (array of size $N$) is divided into two subarrays of half the size, composed of the even and odd elements.
The subarrays are divided in half recursively until there are $N$ leaf arrays consisting of a single element each.
The leaf elements are recombined in the same recursion pattern.
The recombination of subproblems is the addition of the even elements to the result of multiplying the odd elements with the roots of unity (complex-arithmetic operations).

\paragraph{Iterative implementation}
We implement \proj{} using loops to avoid the recursive split of subproblems.
The iterative loop (line 28) starts by recombining the leaf elements where the first half of the input array contains the even elements of the $N$ subproblems, and the second half contains the odd elements.
There are $\log n$ iterations to recombine the even and odd elements of the subproblems.
Every iteration has half the subproblems of twice the size.

\paragraph{Vectorization}
The loops in the code are vectorized by defining array-operand descriptors.
Lines 8-10, as well as lines 32-35, configure the base, length, and stride of array operations.
These are features of the architecture that control the SIMD capabilities of the PEs and remove the overhead of loop management.
Note that because we use complex elements, each element has a real and imaginary part.
Operations over array \texttt{x} are separated for the real and imaginary parts of the complex numbers.
The real-arithmetic equivalent of the complex multiplication with the roots of unity uses four multiplications and two additions.
We combined the additions with multiplications to leverage the native FMAC instructions.

The \texttt{fft task} is performed at each of the $dim$ computation phases (superstep) of a $dim$-dimensional FFT
Each superstep (along a different axis) performs $n^{dim-1}$ pencil computations.

The \proj{} code is configurable at compile time to use single-precision (FP32) or half-precision (FP16).
The storage requirement is $bytes\_per\_complex\_element \times n \times 2$ (two copies of the data are required for the transpose stage between supersteps).
With 48KB of local memory, PEs can handle pencils of up to 4096 complex elements with FP16 or 2048 with FP32.

\subsection{Mapping a 3D FFT into the WSE}\label{sec:3d_mapping_approach}

Figure~\ref{fig:problem_mapping} gives an FFT mapping example with an array of size $4^3$ mapped into a matrix of $4 \times 4$ PEs.
In this figure, and throughout the rest of this section, we describe our largest parallelization possible for a problem of size $n^3$ using $n^2$ PEs, where each PE computes a single pencil per superstep. 
This would be our last datapoint for strong scaling.
Alternatively, each PE could compute several pencils per superstep, as long as the total number of elements fits within the storage limits.
In that sense, a $4^3$ problem could be mapped on a $2 \times 2$ submesh where each PE contains four pencils.
Section~\ref{sec:evaluation} presents the results for both strong and weak scaling.

Figure~\ref{fig:problem_mapping} depicts that, for 3D problems, we choose to start by mapping the x- and y-axis of the input array to the x- and y-axis of the 2D mesh of PEs and the z-axis into memory.
Once the pencil computations along the z-axis have been completed, the \ws needs to perform a data transposition of axes x and z (first transpose) so that the next computation phase computes pencils over the x-axis.
Thus, the first communication phase causes PEs to exchange n-1 (all but one) of their local elements with the other n-1 PEs in their row. 
The second transposition does the same but across PEs in the mesh columns to transpose axes x and y of the 3D array.

We also implemented the inverse FFT (IFFT), which reverses step by step the computation and communication done by the FFT. 
The only difference with IFFT is that the roots of unity have their imaginary part negated and that the result of each compute stage needs to be multiplied by $1/N$.

\subsection{Data Redistribution between Supersteps}\label{sec:data_redistribution_approach}

On the \ws, a PE can propagate one complex element per cycle---as a pair of real and imaginary FP16 values---or every two cycles in FP32 mode.
The bandwidth is critical for the transpose phases since they involve an all-to-all communication of all data elements between the PEs.
The \ws offers explicit, fast communication between the PEs, and that is why it performs well despite having a 2D-mesh, traditionally not ideal for FFT~\cite{mesh_analysis,connection_machine}.

In \proj{}, the redistribution phases are decomposed by each of the $n$ rows or columns of the $n \times $n submesh of the \ws transposing a 2D array.
We will now describe the implementation of the transpose using as an example the first phase (between the PEs of each row).

Picture having a 2D data array of size $n \times n$ initially distributed on a row $P$ of $n$ PEs where each contains a column of data locally, i.e., $x(i,j)$ is on $P(i)$.
Since the goal for each row is to transpose the 2D data matrix split across its PEs, the stream going eastward (starting at PE 0 of the row) transposes elements $x(i,j)$ for $i<j$, i.e., lower triangular submatrix, while the westward stream (starting at PE $n-1$) transposes elements $x(i,j)$ for $i>j$.

Considering $idx$, the index of a PE in the row, the transposition is achieved when the local elements [$0$,$idx-1$] are sent to PEs 0 to $idx-1$ (one element to each PE) and elements [$idx+1$,$n-1$] are sent to PEs $idx+1$ to $n-1$.
Note that these are complex elements, so both real and imaginary components are communicated.

To achieve this communication pattern, we use a \textit{broadcast and filter} approach.
We use two fabric colors to broadcast a data stream in each direction (no link contention), and we program the router filters to capture only one element from each stream, ignoring the rest.
The filters are programmed differently for each router to capture the correct wavelet from the stream. 

Several other methods have been explored based on the \ws routing features, like the point-to-point communication---standard in MPI approaches---but broadcast-and-filter achieved a higher effective bandwidth due to lower router configuration overhead.
A PE takes 30 cycles to set up the routers to distribute its local data.

\subsection{Analyzing a Generalized Redistribution}\label{sec:proof}

Our current transpose implementation only supports using a power of two number of PEs, where each PE contains a single pencil per superstep, but the transpose could be generalized to compute more than one pencil per PE.



We study the \textit{theoretical communication time} ($TT_{comm}$) for 3D FFTs of size $n^3$ on a submesh of $p^2$ PEs where each PE holds $m \times m$ pencils of size $n$, where $m=n/p$.
Since each transpose phase communicates along one dimension of the mesh, and columns or rows communicates in parallel, it is sufficient to study $TT_{comm}$ for a row of PEs.
We denote the PEs of this generic row by $P_0 \ldots P_{p-1}$.
Identical communications occur on each row of the mesh.

We derive $TT_{comm}$ by identifying the most heavily used links (there are two) in the row, counting the wavelets that traverse these links, and dividing by the link bandwidth (one wavelet per cycle).

In the transpose operations, 
$P_i$ sends $p-i-1$ messages towards higher-index PEs and $i$ messages towards lower-index PEs.
Each message is a $m \times m \times m$ cube of complex numbers.

Because of the symmetry, we need only consider the traffic that flows towards higher-index PEs.
We imagine the PEs to be arranged in a line with $P_0$ at the left end and $P_{p-1}$ at the right end.
Each message sent to the right is captured by just one of the router filters, and the other routers just let it flow rightwards.
The data travels all the way to $P_{p-1}$.
Thus, the link leading into $P_{p-1}$ from the left, carries one message from $P_{p-2}$, two messages from $P_{p-3}$, and so on, up to $p-1$ messages from $P_{0}$
Therefore, this link has to transfer $p(p-1)/2$ messages.

In reality, we do not send all of the messages that traverse the last link in one uninterrupted sequence.
There are temporal gaps that add some overhead.
Once a PE has sent all the data it must, it sends a last \textit{control wavelet} to trigger a reconfiguration of the filter in the next router.
There is a delay of $d$ cycles before the next router can start sending data. Since this happens at each router, and there are $p-1$ of them between the left and rightmost, it adds $d\times(p-1)$ cycles to the time.

Therefore $TT_{comm}$ = $p(p-1)/2$ (messages) * $m^3$ (numbers per message) * r (cycles per number) + $d\times(p-1)$ (delay cycles).
Given that we use complex numbers and that the link is 32-bit wide, using FP32 data makes $r=2$, while with FP16 $r=1$.
Since $p=n/m$ and we have measured $d$ to be around 30 cycles, we can consider $TT_{comm}$ as a function of $n$ as:
\begin{equation}\label{eqn:nmf}
TT_{comm(n,m,r)} = n^{2}/2 * m * r - n/2 * m^2 * r + 30(n/m - 1).
\end{equation}
\paragraph{\textbf{Single Pencil:}}
Considering the $m=1$ case, we get:
\begin{equation}\label{eqn:nf}
TT_{comm(n,r)} = n^{2}/2 * r + (30 - r/2)n - 30
\end{equation}
\begin{equation}\label{eqn:fp16}
TT_{commFP16(n)} = n^{2}/2 + 29.5n - 30
\end{equation}
\begin{equation}\label{eqn:fp32}
TT_{commFP32(n)} = n^{2} + 29n - 30
\end{equation}

Since $n^{2} + 29n - 30 <= 2 * (n^{2}/2 + 29.5n - 30)$ for $n\in\mathbb{Z}^{+}$, we conclude that the communication time with FP32 is at most double the time with FP16:
\begin{equation}\label{eqn:fp_ratio}
TT_{commFP32(n)} <= 2 * TT_{commFP16(n)}
\end{equation}

\paragraph{\textbf{Multi Pencil:}}
For the rest of the section we will use the FP32 ($r=2$) case to compare $TT_{comm}$ with different values of $m$.
From the multi-pencil case, we get:
\begin{equation}\label{eqn:multi}
TT_{comm(n,m)} = m * n^2 + (30/m - m^2)n - 30
\end{equation}

Since since $m * n^2 + (30/m - m^2)n - 30 <= m * (n^2 + 29n - 30)$ for $n,m\in\mathbb{Z}^{+}$ and $n>m$, we conclude that the communication time for the multi pencil case is at most $m\times$ the time of single pencil:
\begin{equation}\label{eqn:comm_ratio}
TT_{comm(n,m)} <= m * TT_{comm(n,1)}
\end{equation}

Although not shown here for simplicity, this is also true for FP16.

Cutting the mesh in half ($m=2$) in each dimension while keeping the same problem size leads to almost doubling the communication time.
On the other hand, the compute time goes up by a factor of $m^2$ since the PE has to compute $m^2 \times$ more pencils of the same size $n$.
Therefore, the compute/communication balance improves.

We use these ratios to give an \textit{estimated time} ($ET$) of the $m>1$ cases at
Section~\ref{sec:scaling}, as:
\begin{equation}\label{eqn:expe_comm}
ET_{comm(n,m)} = m * RT_{comm(n,1)}
\end{equation}
\begin{equation}\label{eqn:expe_comp}
ET_{cmpt(n,m)} = m^2 * RT_{cmpt(n,1)}
\end{equation}
where $RT_{comm}$ and $RT_{cmpt}$ are actual CS-2 runtimes of the communication and a computation phases.

\section{Evaluation}\label{sec:evaluation}

We evaluate complex-to-complex fast Fourier transforms for various sizes of three-dimensional arrays and two data precisions.
We report the compute time for in-memory FFT on one PE;
the compute and communication times for 3D FFT with three supersteps and two transposes between the supersteps;
and the weak and strong scaling behavior of our implementation.


\paragraph{Floating-point precision}
We implemented, validated, and evaluated \proj{} using both FP16 and FP32.
The purpose of also evaluating FP16 (less common in the FFT literature) is not to motivate the use of FP16 when computing FFT but to study the performance difference of computing with either precision on the \ws.

\paragraph{Obtaining the runtime}
All results were measured on a CS-2 running at 850 Mhz, having the code and data in the wafer at the start.
The code is compiled using the Cerebras SDK.
We use the hardware cycle counters at each PE to measure runtime.
We report the maximum of the cycle counts of the PEs at the edge of the submesh that we use to run the FFT, as they are the last ones to finish due to the transpose communication.
For 3D problems, we ran forward and inverse Fourier transforms consecutively and reported the runtime divided by two.
We also ran several batches of FFT + IFFT and observed negligible variance ($<1\%$) between runs.

\subsection{In-memory FFT}\label{sec:pencil_compute_results}

From the Cooley-Tukey FFT~\cite{cooley_tukey} formulation that we consider, we know that the real-arithmetic computation of a complex-to-complex FFT of size $n$ contains 5$n \log_2 n$ floating-point operations (flops).

\begin{figure}[t]
\centering  
\includegraphics[width=\columnwidth]{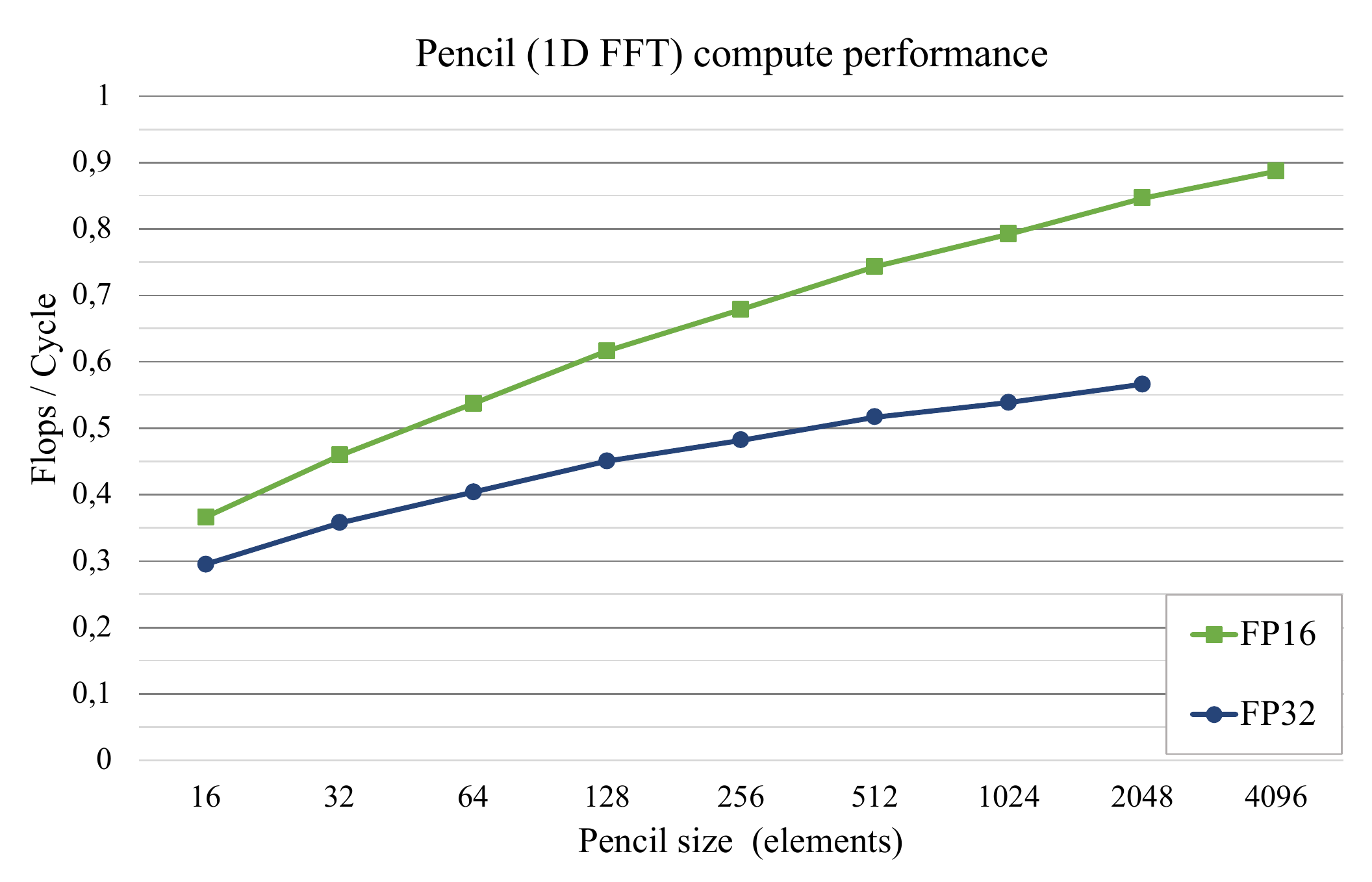}
\vspace{-4mm}
\caption{
Throughput obtained when computing pencils---1D FFTs---of increasing sizes and two precision types on a single PE of the CS-2.
The input and output are in the memory of the PE that performs the computation, and there is no communication with other PEs.
The x-axis is the size $n$ of the pencil being processed.
The y-axis is the throughput in floating-point operations per cycle.
FP16 achieves higher throughput since it has a larger SIMD capability on the WSE.
}
\vspace{-1mm}
\label{fig:pencil_perf}
\end{figure}

Figure~\ref{fig:pencil_perf} shows, for the FP16 and FP32 versions of the \proj{} pencil computation, the values obtained when dividing the 5$n \log_2 n$ flops of computing a 1D FFT on an array of size $n$ by the cycle count of running it on a single PE of the CS-2.
We measure this for power-of-two sizes of $n$, ranging from 16 to 4096.
This last data point is only obtained for FP16, as the PE memory does not fit two arrays (out-of-place computation) of 4096 complex elements in FP32 (each complex element uses 8 bytes).

\paragraph{Effects of SIMD}
We observe that with FP16, a PE obtains a significantly higher throughput than with FP32.
With FP16, a PE performs the vector multiplications and FMAC at lines 38-42 of Listing~\ref{lst:pencil_code} at a rate of four per cycle (SIMD-4), while with FP32, it only performs one per cycle (no SIMD).
Similarly, a PE performs the vector operations of lines 47-49 at a rate of four per cycle (SIMD-4) with FP16, while only two per cycle (SIMD-2) with FP32.
Although the SIMD support raises the throughput, each iteration has a reshape phase (starting at line 52) with runtime $O(n)$ that does not yield flops.

\paragraph{Effects of problem size}
There are $\log_2 n$ iterations of the outer loop starting at line 25 of Listing~\ref{lst:pencil_code}.
Each iteration incurs instruction overheads due to loop management.
Moreover, the vector operations defined in lines 32-35 have a length determined by the variable \texttt{subproblems}.
As we explained in Section~\ref{sec:pencil_compute_approach}, the algorithm starts with $N$ leaf subproblems, and the number of subproblems (and thus the vector size) halves at each outer loop iteration.
With fewer subproblems, each subproblem has more elements and, thus, more iterations of the inner loop (line 28).
The array-descriptor configurations inside the inner loop add $O(n)$ cycles overall.

\paragraph{Overall runtime}
We have looked at the assembly code to analyze the contribution of scalar and vector operations (considering the SIMD of each data type).
We have calculated the cycle count of our implementation to be 3$n \log_2 n$ + 34$n$ + 34$\log_2 n$ with FP16, and 6.5$n \log_2 n$ + 35$n$ + 36$\log_2 n$ with FP32.
(This matches closely the experimental runtime used to obtain Figure~\ref{fig:pencil_perf}.)
We observe that for FP16, the dominant $n \log_2 n$ term is smaller than the number of flops (5$n \log_2 n$) thanks to having SIMD-4.
The $\log_2 n$ terms come from managing the outer loop, while the $n$ terms come from the reshape phases and the inner loop configurations.

The \textbf{takeaway of Figure~\ref{fig:pencil_perf}} is that larger problem sizes better amortize the $n$ and $\log_2 n$ terms.
The last experimental datapoint yields $0.89$ flops per cycle with FP16 and $0.57$ with FP32.
Considering only the $n \log_2 n$ term of our calculation above, the asymptotes are $5/3 = 1.66$ and $5/6.5 = 0.77$ flops per cycle, respectively.


\paragraph{Peak performance}
Extrapolating these results to using the full machine (e.g., to do signal processing or other workload requiring an FFT as part of their process), the peak performance of the CS-2 while doing 850,000 parallel 1D FFTs ($n=2048$) would be 612 Tflops/s with FP16 and 409 Tflops/s with FP32.

\subsection{Multidimensional FFTs}\label{sec:breakdown}

Table~\ref{tab:cycles} shows the cycle counts for the different 3D problem sizes and precision that we measured on the CS-2.
We evaluated 3D FFTs of size $n^3$, for $n$ every power of two from 32 to 512, on a subset of a CS-2 mesh of size $n \times n$.
Therefore, we employ up to $512 \times 512$ PEs, which is about 70\% of the PEs of a CS-1 and only 30\% of a CS-2. 
We are not utilizing the entire fabric on the CS-2 because \proj{} computes 3D arrays in which every dimension has power-of-two elements, using power-of-two PEs per dimension of the CS-2.

Our 3D FFT starts with the input data distributed to the mesh, where each processor holds a pencil.
Then each processor computes the FFT of its pencil, and every row of PEs performs a matrix transpose and computes again.
Then, another transpose happens on every column of PEs before the last computation phase.
So there are a total of three one-dimensional parallel Fourier transform steps and two redistributions (one on each axis of the mesh).
We measured the cycles from the computation and communication phases separately to showcase their contribution to the total runtime.

\begin{figure}[t]
\centering  
\includegraphics[width=\columnwidth]{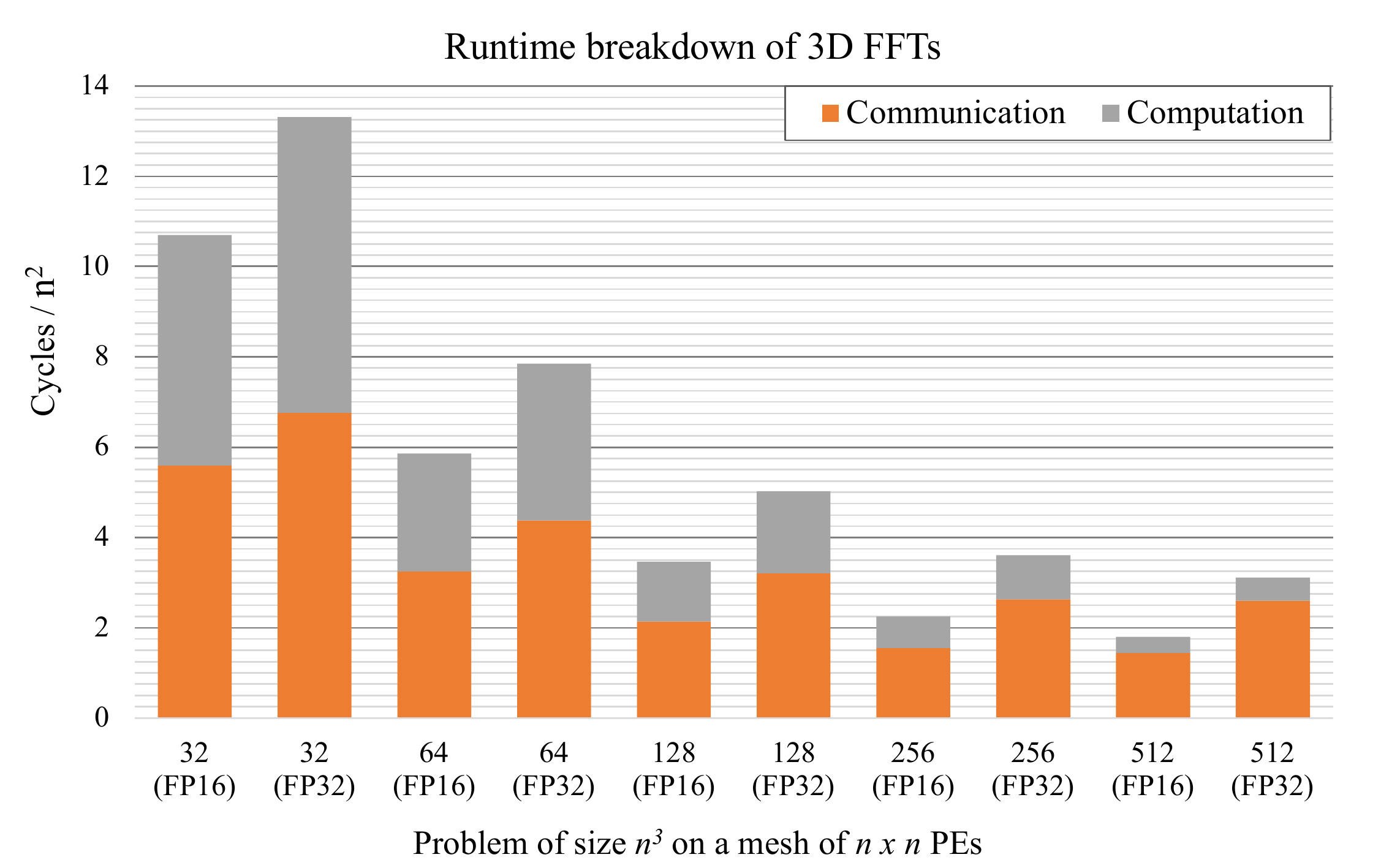}
\vspace{-2mm}
\caption{
Breakdown into communication and computation of the runtime of running 3D FFT for the problem sizes and CS-2 configurations reported in Table~\ref{tab:cycles}.
The x-axis indicates $n$.
The y-axis indicates cycles divided by $n^2$ to better analyze the communication time.
}
\vspace{-2mm}
\label{fig:breakdown}
\end{figure}

Figure~\ref{fig:breakdown} breaks down the runtime reported in Table~\ref{tab:cycles} into these phases.
The cycle count is divided by $n^2$ to obtain the constant factor of \proj{} implementation concerning the $O(n^2)$ communication time.
In this weak scaling study, we grow the \ws's subgrid with the problem size, so that each PE only computes one pencil per superstep.
This means that the computation time remains as $O(n \log_2 n$) and represents a smaller part of the runtime as $n$ grows.

Since we have two transpose phases on a 3D FFT, the asymptote of the communication time is $n^2$ for FP16 and $2 \ n^2$ for FP32.
We observe from Figure~\ref{fig:breakdown} that the $O(n)$ router configuration overheads that we calculated in Equations \ref{eqn:fp16} and \ref{eqn:fp32} become less significant when running larger problem sizes.
Thus, the cycle count gets closer to the asymptotic $O(n^2)$ terms.

\begin{table}
  \caption{Cycle counts measured on the CS-2 when running 3D FFTs with increasing problem and mesh sizes, using FP16 and FP32.}
  \label{tab:cycles}
  \begin{tabular}{cccc}
    \toprule
    Problem Size & Submesh Size & Cycles FP16 & Cycles FP32 \\
    \midrule
    $32^3$  & 32 $\times$ 32 & 10,953 & 13,633 \\
    $64^3$  & 64 $\times$ 64 & 24,000 & 32,176\\
    $128^3$ & 128 $\times$ 128 & 56,741 & 82,405\\
    $256^3$ & 256 $\times$ 256 & 147,247 & 236,329\\
    $512^3$ & 512 $\times$ 512 & 471,064 & 815,371\\
  \bottomrule
\end{tabular}
\end{table}

\subsection{Scaling of \proj{} up to a Million Cores}\label{sec:scaling}

\begin{figure}[t]
\centering  
\includegraphics[width=\columnwidth]{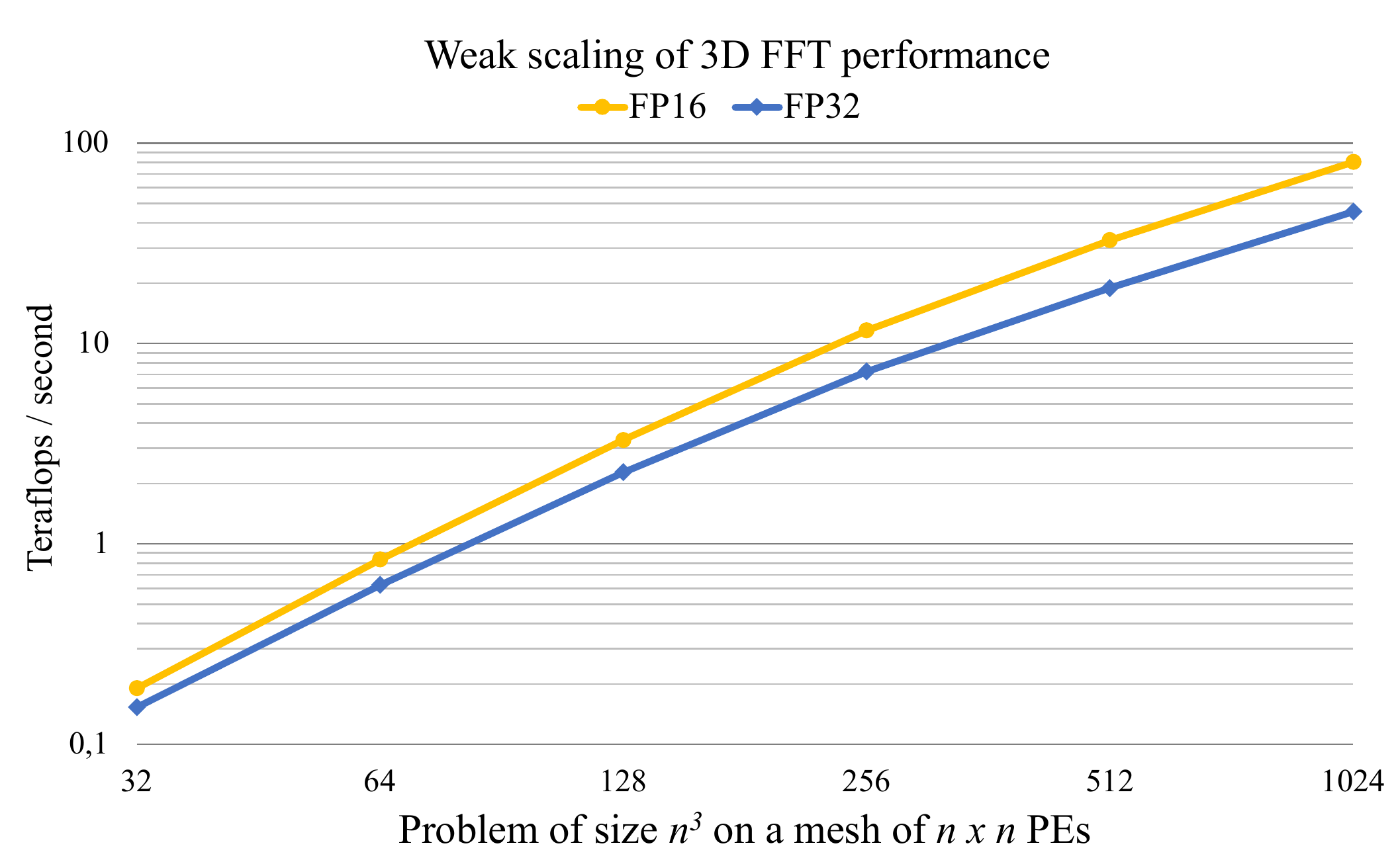}
\vspace{-4mm}
\caption{
Weak scaling performance, in teraflops per second, when performing 3D FFT on $n^3$ elements, mapped into $n \times n$ PEs. 
The x-axis indicates $n$.
}
\vspace{-1mm}
\label{fig:weak_scale}
\end{figure}

\begin{figure}[t]
\centering  
\includegraphics[width=\columnwidth]{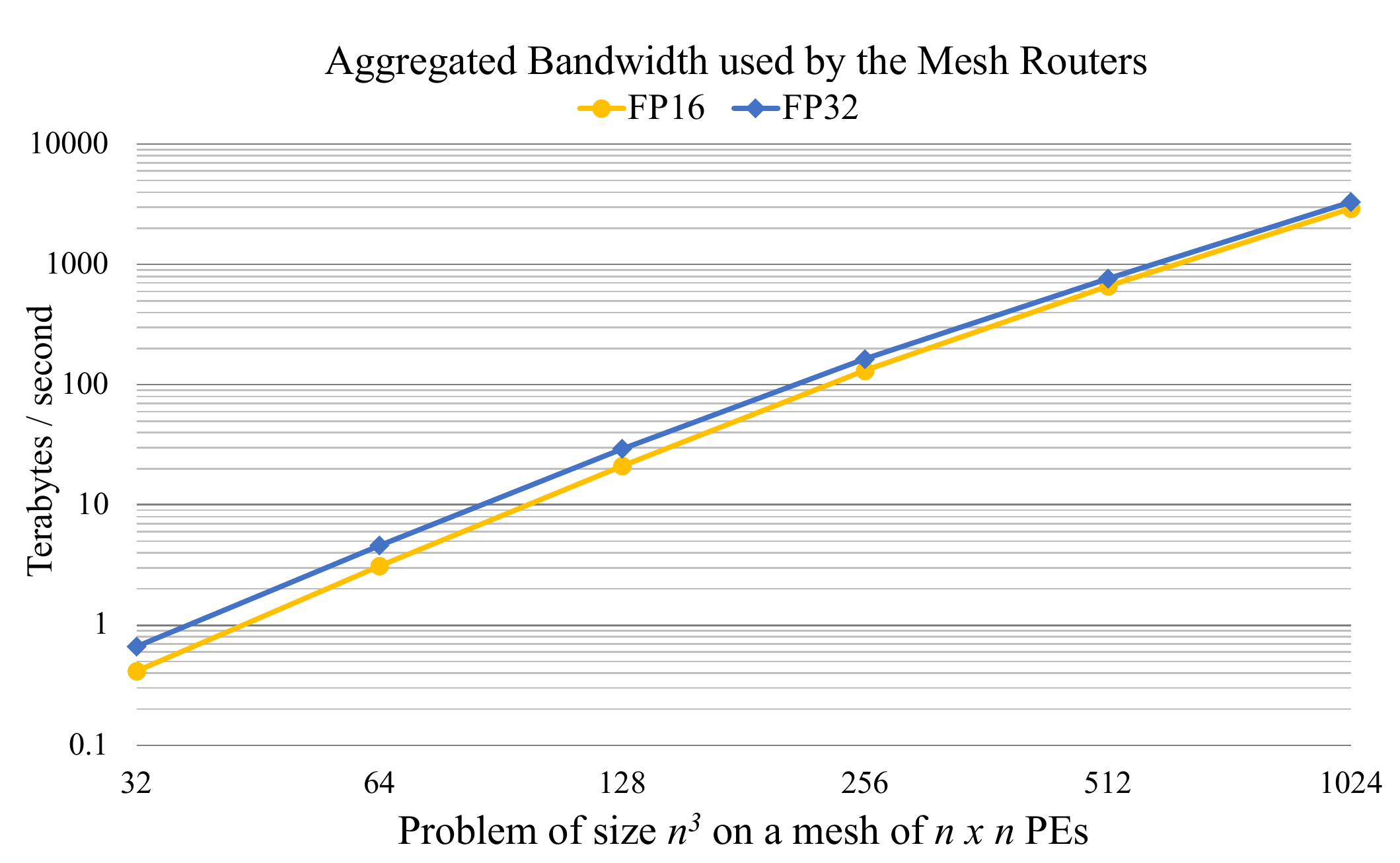}
\vspace{-4mm}
\caption{
Bandwidth utilized by the mesh routers, in terabytes per second, when performing 3D FFT on $n^3$ elements, mapped into $n \times n$ PEs. 
The x-axis indicates $n$.
}
\vspace{-2mm}
\label{fig:weak_scale_bw}
\end{figure}

This section combines the CS-2 measurements shown in Table~\ref{tab:cycles}, with some lower-bound throughput estimates of other problem mappings to show weak and strong scaling of \proj{}.

\paragraph{Weak scaling:}
Figures~\ref{fig:weak_scale} and \ref{fig:weak_scale_bw} shows the performance and network bandwidth of \proj{} with problems of size $n^3$, mapped into $n \times n$ PEs, for several power-of-two values of $n$.
For all problem sizes, there is a single pencil per PE.
Figure~\ref{fig:weak_scale} values are obtained dividing the number of floating-point operations performed ($3 \ n^{2} \times 5 n \log_2 n$) by the runtime; while
Figure~\ref{fig:weak_scale_bw} values come from dividing the number of bytes routed by every router of the submesh involved in the 3D FFT, divided by the total runtime.
The runtime is obtained from the cycles of Table~\ref{tab:cycles} at 850Mhz, except for the last problem size ($n=1024$), which is estimated using a possible hypothetical machine with 1024 PEs per dimension (a bit over a million cores).
Given that the CS-2 already has around 850,000 PEs, and this amount doubled since the previous generation, such a machine might be possible in the future, and we already provide a precise estimation of what the performance of \proj{} would be.
We estimate the runtime of the $1024^3$ case as:
\begin{equation}\label{eqn:exp2}
ET_{total(n=1024)} = 4 * RT_{comm(n=512)} + 3 * RT_{pencil(n=1024)}
\end{equation}
We had already measured $RT_{pencil(n=1024)}$ while creating Figure~\ref{fig:pencil_perf}.
Since all the PEs do this in parallel, the whole computation superstep takes this same time, and there are three such supersteps in a 3D FFT separated by data redistributions.
We make an upper bound estimation of the total communication time as $4 * RT_{comm(n=512)}$ since using Equation~\ref{eqn:nf} we conclude that $RT_{comm(2n)} <= 4 * RT_{comm(n)}$.

\paragraph{FP32 and FP16 scaling:}
Regarding actual measurements, for $n=512$ we are achieving $18.9$ Tflops/s with FP32, and $32.7$ Tflops/s with FP16.
For this problem size and data mapping ($m=1$), the communication is dominant for both data types.
From Equation~\ref{eqn:fp_ratio} we know that the communication time with FP32 is at most double the time with FP16.
However, we are not getting half the Tflops/s with FP32 because---as we observed from the breakdown in Figure~\ref{fig:breakdown}---the communication time with FP32 for $n=512$ is $1.8\times$ longer.
This is due to the $O(n)$ overhead of the route reconfiguration, which has a more significant impact with FP16 (Equation~\ref{eqn:fp16}).

\paragraph{Network bandwidth:}
Figure~\ref{fig:weak_scale_bw} showcases that FFT is  as much about moving data as it is about flops, and parallelizing over such a large processor grid requires extremely high network bandwidth.
The transpose implementation that we described in Section~\ref{sec:data_redistribution_approach} requires that each PE sends $n-1$ data to the other $n-1$ PEs.
Since the PEs are distributed in a mesh, this means that the data may transit many hops to reach its destination.
At 4 bytes per hop, we get that the total bandwidth utilized by the $512\times512$ routers involved in computing a $512^3$ FFT is 0.8 Petabytes/s.

\paragraph{Problem sizes:}
Looking at the other problem sizes in these scaling plots, although $32^3$ is a small problem, it is an interesting size to run in the CS-2 when many of them are done in parallel; for example, in one of the domains in which FFT is a key component, or even as a decomposition of a higher dimensional FFT.
These sizes evaluated are not arbitrary since we find that 3D FFT of $n$ = 256, 512, and 1024 are pretty important for HPC applications in chemistry and cosmology.
These are also found in the evaluation of prior FFT literature.
This is due to the use of FFT in the time-stepping loop in molecular dynamics codes (like LAMMPS~\cite{lammps}) and cosmology codes (like HACC~\cite{Habib_2016}), for example, where the time taken per timestep is critical, given the millions of timesteps in a simulation.
Thus, it is interesting to look at strong scalability.

\begin{figure}[t]
\centering  
\includegraphics[width=\columnwidth]{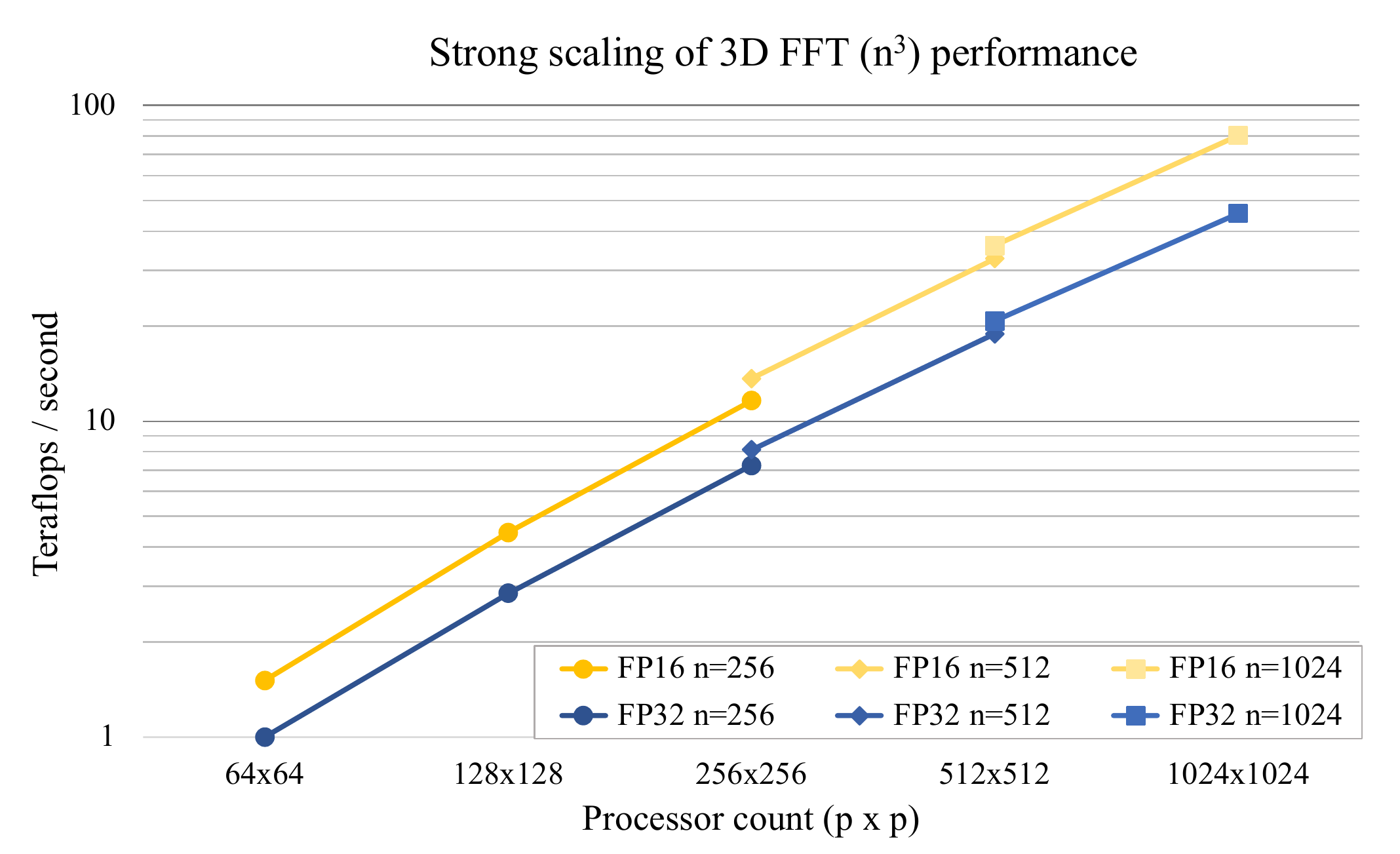}
\vspace{-4mm}
\caption{
Strong scaling performance (in Tflops/s) of FFTs of size $256^3$ (on a $64\times64*$, $128\times128*$, and $256\times256$ mesh), $512^3$ (on a $256\times256*$, and $512\times512$ mesh) and $1024^3$ (on a $512\times512*$, and $1024\times1024*$ mesh).
The datapoints marked with $*$ are estimations, and the rest are actual CS-2 runs.
}
\vspace{-3mm}
\label{fig:strong_scale}
\end{figure}

\paragraph{Strong scaling:}
Figure~\ref{fig:strong_scale} shows the 3D FFT performance of three problem sizes ($256^3$, $512^3$ and $1024^3$) with increasing square-shaped subsets of PEs.
For a given $n$, each strong scaling step involves doubling the number of PEs per dimension ($p$) until $n=p$.
The last datapoint for a given problem size represents a mapping with as many PEs per dimension ($p$) as $n$.
Since $m=n/p$, in the last datapoint for a given problem size, each PE computes a single pencil per superstep ($m=1$).
We already showed the cycle counts for the single-pencil case in Table~\ref{tab:cycles}.
In preceding datapoints, each PE contains $m^2\times$ more pencils (of the same size $n$), since $p$ halves and thus $m$ doubles.
Regarding data redistribution, since the total data to communicate remains the same while the bisection bandwidth halves, the communication time would, in theory, double.
As we concluded in Equation~\ref{eqn:comm_ratio}, this is an upper-bound estimation because each row/column of PEs communicates twice the data but over fewer PEs---causing less route reconfiguration overhead.

We use Equations \ref{eqn:expe_comm} and \ref{eqn:expe_comp} to estimate the total runtime of datapoints with $m>1$, as:
\begin{equation}\label{eqn:expe3}
ET_{total(n,m)} = m * RT_{comm(n,1)} + m^2 * RT_{cmpt(n,1)}
\end{equation}

Both $RT_{comm}$ and $RT_{cmpt}$ are the aggregated time spent on the CS-2 in the communication and computation phases. We use the same data as in the runtime breakdown shown in Figure~\ref{fig:breakdown}.

Strong scaling on the \ws gets speedups from two sources: each PE computes fewer pencils and the bisection bandwidth doubles.
Doubling $p$ (and thus halving $m$) approximately results in: 
\begin{equation}\label{eqn:expee}
ET_{total(n,m/2)} = RT_{comm(n,m)} / 2 + RT_{cmpt(n,m) / 4}
\end{equation}
Since the total number of PEs grows by four at each step, there is a perfect reduction in $RT_{cmpt}$ but not in $RT_{comm}$.
For example, we observe a 2.85$\times$ speedup when scaling FP32 $256^3$ from $64\times64$ to $128\times128$, and a 2.54$\times$ speedup on the next scaling step.
Although not perfect scaling, these speedups are good considering the fact that we are parallelizing $n^3$ problems with up to $n^2$ processors, where messages between processors are only 8-byte long.

\subsection{Comparing with State-of-the-art Machines}\label{sec:comparison}

By way of comparison, we looked at the fastest FFT codes running on state-of-the-art machines.
Table~\ref{tab:competition} compared their reported Tflops/s to ours for the same problems.

For example, using single-precision, a 3D FFT problem of size $n=512$ is reported to run at 16 Tflops/s with the Cuda-written cuFFT on the latest DGX of Nvidia GPUs~\cite{nvidia_fft}.
This performance is identical on a 16-GPU V100 DGX2 unit and the newer 8-GPU A100 DGX unit.
The measured performance of \proj{} in the CS-2 for that same precision and size is 18.9 Tflops/s, which is 18\% better than any DGX performance.

For a single-precision, $n=1024$, 3D FFT, Nvidia reports 20 Tflops/s on their DGX2 cluster and around 19 Tflops/s on the A100 DGX.
As we showed in Figure~\ref{fig:strong_scale}, with that precision, we would obtain 22.5 Tflops/s on a $512 \times 512$ submesh of the current CS-2. With half-precision, the performance is 36 Tflops/s.
A possibly future mesh with 1024 PEs per dimension would achieve 49.7 Tflops/s using single-precision and 80 Tflops/s using half-precision.

On this $1024^3$ problem, Google reports~\cite{google_fft} 14.8 ms, that is, 10.9 Tflops/s (single-precision), using a TPU v3 pod with 2048 cores.

\paragraph{Can one use larger machines to accelerate FFT without running enormous FFT sizes?}
Recently, experiments have been done on a very large cluster, the Oak Ridge National Laboratory supercomputer, Summit~\cite{summit}, which has over 4500 compute nodes, each with two IBM Power9 CPUs and six V100 GPUs.
Summit has a powerful Infiniband interconnect with maximum bisection bandwidth provided by an untapped fat tree.
So, in theory, it should be fast.
In practice, a study~\cite{heffte_report} by a team at the University of Tennessee measured up to 9 Tflops/s for $1024^3$ 3D FFT with double-precision complex data on 1024 nodes of Summit (6144 GPUs).
They found that the data redistribution phases take over 90\% of the time on Summit.
It appears that the message granularity for a $1024^3$ problem mapped across such a large machine produces poor interconnect performance, likely due to the message latency rather than the bisection bandwidth.
The Summit's scaling steps, from 256 to 512 nodes, and 512 to 1024, yield less than 50\% performance improvement.

\paragraph{Strong scaling leads to small messages between cores}
The computation of FFTs, even of a billion elements, is not a suitable problem for acceleration on supercomputers.
Due to the inefficiencies of small message sizes in MPI communication, strong scaling fails, and even a vast machine like Summit cannot do better than \proj{} for the problem sizes evaluated.

\begin{table}
  \caption{Comparing \proj{} with the fastest reported FFT times for different problem sizes. The $1024^3$ datapoints marked with * are estimations of \proj{} on a $512 \times 512$ mesh.}
  \label{tab:competition}
  \begin{tabular}{cccl}
    \toprule
    Size & Precision & Software \& Machine & Tflops/s\\
    \midrule
    $256^3$ & 64-bit & Takahashi~\cite{takahashi2009_cluster} on Appro Xtreme-X3 & 0.4 \\
    $256^3$ & 64-bit & HeFFTe~\cite{heffte} on 32-node Summit & 0.5 \\
    $256^3$ & 32-bit & \proj{} on CS-2 & 7.2 \\
    $256^3$ & 16-bit & \proj{} on CS-2 & 11.6 \\
    \hline
    $512^3$ & 64-bit & HeFFTe~\cite{heffte} on 64-node Summit & 1.3 \\
    $512^3$ & 32-bit & cuFFT~\cite{nvidia_fft} on DGXA100 & $\sim$16 \\
    $512^3$ & 32-bit & \proj{} on CS-2 & 18.9 \\
    $512^3$ & 16-bit & \proj{} on CS-2 & 32.7 \\
    \hline
    $1024^3$ & 64-bit & HeFFTe~\cite{heffte} on 1024-node Summit & $\sim$9 \\
    $1024^3$ & 32-bit & Google's FFT on TPU v3 pod~\cite{google_fft} & 10.9 \\
    $1024^3$ & 32-bit & cuFFT~\cite{nvidia_fft} on DGXA100 & $\sim$19 \\
    $1024^3$ & 32-bit & \proj{} on CS-2* & 22.5 \\
    $1024^3$ & 16-bit & \proj{} on CS-2* & 36 \\
  \bottomrule
\end{tabular}
\vspace{-2mm}
\end{table}

\section{The \ws and the mesh for All-to-all Communication}\label{sec:all-to-all}

We have seen that \proj{} mapping of a 3D problem into the mesh, as a 2D FFT per row or column, leads to a transpose time of $n^2$.
Although we have not evaluated a 2D FFT mapped into a rectangle of PEs of the CS-2, we can get a good idea of its communication time on the \ws, from a lower bound due to the bisection bandwidth.
Both the bound and the actual communication time greatly depend on the networking topology of the machine.

\subsection{Communication Time of 2D Problems}

Let us assume a $n^2$ problem distributed on $n$ processors interconnected a mesh of $\sqrt{n} \times \sqrt{n}$ PEs.
Considering the bisection of this network, there are $\sqrt{n}$ communication links connecting the two halves due to the simple mesh interconnection topology.
And since all processors send an equal amount of their data, in doing the transpose, to each other processor, we see that half of the data on the left half is sent to the right half and vice versa.
Thus, $n^2 / 4$ elements must transit the $\sqrt{n}$ links from left to right and the same amount in the other direction.
Given that the \ws has a bidirectional link and the bandwidth is one word per clock, it would take $n^2 / 4 / \sqrt{n}$ cycles to get that data across with FP16 and twice as much for FP32.
That should be compared to the number of floating-point operations that each processor does for a one-dimensional transform of a vector of length $n$, which is $O(n \log_2 n)$ time.

Although the midline of the mesh is a communication bottleneck, a small problem size would not show it entirely.
The communication time only becomes dominant with very large meshes of PEs.

When transforming an image of size $n^2$ on a mesh of $\sqrt{n} \times \sqrt{n}$ PEs, the parallel computation time scales as $n \log n$, and the communication grows as $n \sqrt{n}$.

\subsection{Topology Considerations}

Meshes are not a popular interconnection topology in supercomputers.
Designs for such systems prefer costlier networks with greater bandwidth across bisections of the machine; the “fat tree” is such a network.
A fully configured fat tree has bisection bandwidth that scales linearly with the number of computation nodes (a node is a server in the cluster).
In contrast, the bisection bandwidth in a mesh is only the square root of the processor count.
Although the \ws has less bisection bandwidth, it achieves comparable performance, and the reasons are quite fascinating.

\paragraph{The \ws mesh:}
Rather than meter-scale inter-node cables, the \ws processors communicate via sub-millimeter scale wires.
These wires are dense, allowing huge network bandwidth while consuming much less energy---well under a picojoule to send a bit.
On the \ws, each PE can communicate with its neighbors as fast as it can compute.
A one-word message can be sent on one machine cycle and arrive at the neighboring processor on the next cycle.
This is three orders of magnitude better than the latency for small messages in clusters.
When an application communicates in a mesh-like pattern---typical for finite element and finite difference methods---the WSE mesh interconnect provides enough bandwidth that fine-grained parallelism works quite well, as a study previous study by the National Energy Technology Lab showed~\cite{cerebras_stencil}.
But FFT is different, the all-to-all communication leads to a lot of small messages between distant processors.

\paragraph{Bisection bandwidth:}
For a $512 \times 512$ mesh, we can send 512 four-byte words per clock across the bisection, in each direction, and at 850 MHz which equates to 3.5 TB/s of bisection bandwidth.
Supercomputers that take three orders of magnitude more power have between one and two orders of magnitude more bisection bandwidth.
For instance, Summit, a machine roughly 100 times the size and cost, has about 115 TB/s of bisection bandwidth~\cite{summit_next}.
Summit has a full fat-tree topology, giving it a bisection bandwidth equal to its injection bandwidth, which is typical in supercomputers.

\paragraph{Fine-grain communication:}
Despite the bandwidth advantage, we have seen that the \ws achieves comparable performance on FFTs of common, relevant sizes. 
The reason is that, on the one hand, clusters pay an MPI message overhead for inter-node communication and longer latencies, which becomes a burden with the fine-grain communication that arises from large parallelizations (as discussed in Section~\ref{sec:comparison}).
On the other hand, the \ws has all the computing elements on-chip, and even the communication of single-word messages has minimal overheads, as shown in this paper.

\section{Additional Related Work}\label{sec:related}


In the world of distributed-CPU libraries, FFTW~\cite{fftw} supports MPI via slab decomposition, and so  it is limited to running a small number of nodes. 
P3DFFT~\cite{p3dfft} extends FFTW and supports pencil decompositions. 
Some HPC applications such as LAMMPS~\cite{lammps}---for molecular-dynamics simulations---have built their own FFT library, like fftMPI~\cite{plimpton}, which improved using prior implementations by using new MPI features for one-sided communication~\cite{one_sided}.
Dalcin et al.~\cite{dalcin_advanced_mpi} also leveraged advanced features like the generalized all-to-all scatter/gather from MPI-2 to communicate dis-contiguous memory buffers, thus eliminating the need for local data realignments. Dalcin also provided a Python binding with mpi4py-fft~\cite{fftmpi_py}.

Our implementation, \proj{}, does not use MPI libraries since communication is built into the Cerebras instruction set.
The \ws allows similar communication patterns as modern MPI while providing better performance since communication happens on-chip and has little message overhead.

Regarding the FFT computation itself, there are many implementations in the literature, targeting all kinds of hardware ~\cite{fftw, fftpack, mkl, nvidia_fft,google_fft}. 
HeFFTe~\cite{heffte} agglomerates several of them by having C++ and Cuda kernels for computation and MPI and OpenMP interfaces for communication.
Since Cerebras has its own architecture, \proj{} is made in the Cerebras language using their new SDK.
Our implementation is based on the Cooley-Tukey algorithm~\cite{cooley_tukey} and is optimized for the \ws's SIMD capabilities. An open question is whether another algorithm for FFT than Cooley-Tukey would perform better in the \ws.

In the previous section, we have shown how \proj{} compares with fastest FFT implementations on state-of-the-art hardware: cufft on DGX~\cite{nvidia_fft}, Google's implementation on a TPU cluster~\cite{google_fft} and HeFFTe~\cite{heffte} on Summit~\cite{heffte_report}.
We have also looked at but not shown comparisons with other CPU implementations~\cite{fft_on_mira,plimpton,p3dfft,fftw}, due to having been outperformed by the previously mentioned approaches.

\section{Discussion: Larger FFT Problems}\label{sec:discussion}


In this paper, we have studied the communication time of data redistribution phases of 2D and 3D FFTs in a wafer-scale mesh and evaluated the last ones on the CS-2.
We also evaluated the computation time of the Fourier transform in a local PE.
Although we have not evaluated very large one-dimensional arrays, like the ones used for signal processing, these could be decomposed as a higher dimensional FFT.

We have studied problem sizes within the memory capacity of the \ws, up to $1024^3$.
Problems whose memory footprint does not fit in the wafer could be solved by decomposing the work into several lower-dimensional problems.
For example, a 3D FFT can be decomposed into several 2D problems or into many more one-dimensional problems. 
Data would be streamed in and out of the fabric, and the PEs are processing subproblems as they come.
However, the communication overhead of such an approach might not make it competitive.

Nonetheless, the range of problems whose data footprint is within the wafer storage is relevant in several domains, where FFTs would occur in between application phases, and the data is already in place.
Examples of this include image analysis, tomography, and partial differential equations.
Alternatively, if the dataset for those applications was distributed across several instances of the \ws, and an FFT was needed with the data in place, it could follow the same scheme of \proj{}, where the transpose would happen across systems.


\section{Conclusion}\label{sec:conclusions}

Fast parallel FFT remains crucial due to the central importance of FFT in HPC grand challenge applications, including in chemistry, cosmology, and other situations.
Often, the desire is to accelerate a moderately large 3D FFT via strong scaling.
That is a daunting challenge for supercomputers, which have poor small-message communication characteristics and may have limited bisection bandwidth.

This paper examined the weak and strong scaling capability of wafer-scale processing for problems of this kind, focusing on the 3D FFT of size $512^3$, a typical model problem.
We have used the only wafer-scale systems available today, the Cerebras systems.
These employ a wafer-scale engine (\ws) containing all processors, memory, and interconnection network hardware.

We have measured time using up to 262,144 cores (a $512^2$ processor mesh), which is roughly one-third of the CS-2, and estimated the performance of \proj{} on a possible future mesh containing over a million cores ($1024^2$). 
The Cerebras \ws, it turns out, is a very effective way to make the FFT algorithm run even faster than has been possible until now.
As fast as a tens-of-megawatts supercomputer, but on a single silicon piece.
This is so because FFT demands more communication than computing.

The main question with FFT is whether the data redistribution between the unidirectional FFT phases renders the overall runtime competitive.
We have found that the transposes dominate the runtime on the \ws---up to 80\% for problem sizes of interest---even though the experimental results achieve a nearly theoretical communication time on a 2D mesh ($n^2$).
The communication dominates the runtime because the computation is parallelized to the extreme---each PE computing a single pencil---and so the whole compute superstep takes $O(n \log n)$ cycles.

The net result is a machine runtime of 959 microseconds for the $512^3$ three-dimensional FFT, which is the fastest achieved for this benchmark in the current literature.





\bibliographystyle{ACM-Reference-Format}
\bibliography{refs}
\end{document}